\documentclass[useAMS,usenatbib]{mn2e}
\usepackage[dvips]{graphicx}
\usepackage{psfig}

 \title[An M9.0 dwarf at 8~pc]{Discovery of a nearby M9 dwarf}
 \author[N. Phan-Bao et al.]{N. Phan-Bao$^{1}$\thanks{E-mail: 
              pbngoc@asiaa.sinica.edu.tw (PBN)},
  M.S. Bessell$^{2}$, E.L. Mart\'{\i}n$^{3,4}$, G. Simon$^{5}$, 
  J. Guibert$^{6}$, T. Forveille$^{7,8}$, 
  \newauthor 
  X. Delfosse$^{8}$, F. Crifo$^{5}$, N. Epchtein$^{9}$, P. Wood$^{2}$, 
  F. Tajahmady$^{6}$\\
$^{1}$Institute of Astronomy and Astrophysics, Academia Sinica.
           P.O. Box 23-141, Taipei 106, Taiwan, R.O.C. \\
$^{2}$Research School of Astronomy and Astrophysics, Australian National 
           University, Cotter Rd, Weston, ACT 2611, Australia. \\
$^{3}$Instituto de Astrof\'{\i}sica de Canarias, C/ V\'{\i}a L\'actea  
      s/n, E-38200 La Laguna (Tenerife), Spain. \\
$^{4}$University of Central Florida, Dept. of Physics, PO Box 162385, 
          Orlando, FL 32816-2385, USA. \\
$^{5}$GEPI, Observatoire de Paris, 5 place J. Janssen, 92195 
      Meudon Cedex, France. \\
$^{6}$Centre d'Analyse des Images, GEPI, Observatoire de Paris,  
      61 avenue de l'Observatoire, 75014 Paris, France. \\
$^{7}$Canada-France-Hawaii Telescope Corporation, 65-1238 Mamalahoa 
      Highway, Kamuela, HI 96743 USA.	 \\
$^{8}$Laboratoire d'Astrophysique de Grenoble, Universit\'e J. 
          Fourier, B.P. 53, F-38041 Grenoble, France. \\
$^{9}$LUAN/UNSA/UMR-CNRS 6525, Parc Valrose, F 06108 NICE Cedex 2, France.}
\begin{document}
      \date{Received / Accepted}
\pagerange{\pageref{firstpage}--\pageref{lastpage}} \pubyear{2005}      

\maketitle

\label{firstpage}

\begin{abstract}
We report the discovery of a new M9.0 dwarf at only 8.2~pc, which we 
identified in our search for nearby ultracool dwarf ($I-J \geq 3.0$, 
later than M8.0) in the DENIS database. We measure a very high proper 
motion of 2.5~arc-sec/yr. The PC3 index measured from its low-resolution 
spectrum  gives a spectrophotometric distance of 8.2~pc. This makes it the 
third closest M9.0 dwarf.
\end{abstract}



\section{Introduction}

Nearby stars are the brightest representatives of their class,
and therefore provide benchmarks for stellar physics. This is 
particularly true for intrinsically faint objects, such as white dwarfs, 
stars at the bottom of the main sequence, and brown dwarfs (BDs). In the 
last decade a significant number of nearby ultracool dwarfs have been 
identified by the DENIS \citep{epchtein97} and 2MASS \citep{skrutskie}
surveys. On the DENIS side \citet{delfosse01} reported an 
M9 dwarf at 5~pc,  and \citet{martin99} found a late-L dwarf at the same
distance (DENIS-P J0255$-$4700, \citealt{martin99}; \citealt{cruz03}).
Based on the 2MASS survey \citet{burgasser00} discovered a T5 dwarf at 10~pc
(2MASS 0559$-$14, \citealt{dahn02}), and \citet{teegarden} recently 
reported an M6.5 at only $\sim$4~pc. These very close ultracool dwarfs 
are much brighter than more distant objects and clearly easier to observe.

We are mining the DENIS database for nearby ultracool dwarfs, and have 
reported our intermediate results in several publications (e.g. 
\citealt{delfosse97}, \citealt{martin99}, \citealt{phan-bao01, phan-bao03}). 
Our search has a limiting distance of $\sim$30~pc, but ultracool 
dwarfs within 10~pc are obviously of particular interest. Here
we report our detection of a M9.0 dwarf at only 8~pc, 
LEHPM 3396 or DENIS-P J033411.39$-$495333.6 (hereafter DENIS 0334$-$49) 

Section~2 decribes the observational data and their analysis, and
Section~3 discusses our spectral type and distance estimates.

\section{Observational data}
\begin{figure}
\psfig{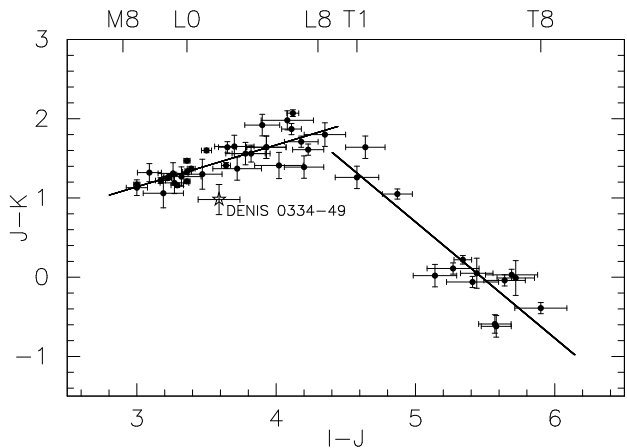}
\caption{$I_{\rm C}-J$, $J-K_{\rm S}$ diagram for 48 ultracool dwarfs 
($I_{\rm C}-J \geq 3.0$); most photometric
data from \citet{phan-bao03}, \citet{dahn02}, \citet{reid01},
\citet{burgasser03} and references therein. The distinct M/L and T 
sequences are clearly visible.}
\label{fig_sequence}
\end{figure}

\begin{figure}
\psfig{width=8.5cm,file=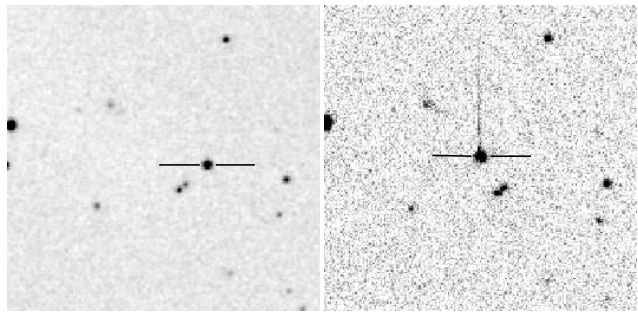,angle=0}
\caption{Archival images of DENIS 0334$-$49: SERC-I ({\it left}, epoch: 
1984.915) and DENIS-I ({\it right}, epoch: 1996.937). The size of each 
image is 3.2\arcmin$\times$3.2\arcmin, and North is up and East to the left.}
\label{fig_pm}
\end{figure}

\begin{table*}
   \caption{Coordinates at the DENIS epoch, photometry and astrometry of DENIS 0334$-$49}
    \label{astrometry}
  $$
   \begin{tabular}{lllllllllll}
   \hline 
   \hline
   \noalign{\smallskip}
$\alpha$(2000.0) & $\delta$(2000.0)  & Epoch &  B  &  R  &  I  &  J  &  K & $\mu_{\alpha}$(\arcsec/yr) & $\mu_{\delta}$(\arcsec/yr) & 
$\mu_{\rm total}$(\arcsec/yr) \\
    \noalign{\smallskip}
    \hline 
03 34 11.39 & $-$49 53 33.6 & 1996.937  &  20.5    &  17.5    &  14.90 & 11.31  & 10.33 & 2.35 & 0.47 & 2.50$^{a}$   \\
            &               &           & $\pm$0.3 & $\pm$0.2 & $\pm$0.05 & $\pm$0.10 & $\pm$0.09 & $\pm$0.03 & $\pm$0.03 & $\pm$0.03 \\
   \noalign{\smallskip}
    \hline 
   \end{tabular}
  $$
  \begin{list}{}{}
  \item[$^{\rm a}$]: previously known as a high proper motion star, as 
      LEHPM 3396 in \citet{pokorny}
  \end{list}  
\end{table*}
We selected DENIS 0334$-$49 in the DENIS database from its colours, which
fall in the area of the ($I-J$, $J-K$) diagram occupied by very late-M and 
L dwarfs (Fig.~\ref{fig_sequence}). Fig.~\ref{fig_pm} shows its finding chart.
We then proceeded to measure its 
proper motion, to discriminate against a distant giant of similar colour. 
DENIS 0334$-$49 appears on 4 plates in the collection of the Centre 
d'Analyse des Images (CAI, {\footnotesize http://www.cai-mama.obspm.fr/}):
SERC-I, SERC-J, SERC-R and ESO-R. We digitized these survey plates with
the MAMA microdensitometer \citet{berger} at CAI, and analysed the 
resulting images with SExtractor \citep{bertin}. We calibrated these 
measurements using the ACT \citep*{urban} catalogue \citep{postman}
as astrometric reference and the GSPC-2  \citep{bucciarelli}) catalogue
as photometric reference. A least-square fit to these 4 positions and
the DENIS and 2MASS positions determines an absolute proper motion
of 2.5~arc-sec/yr, over a baseline of 22~years. At this point we realized 
that DENIS 0334$-$49 had been previously identified as a high proper
motion star by \citet{pokorny}, under LEHPM 3396. 
Table \ref{astrometry} summarizes the photometry and astrometry of 
DENIS~0334$-$49. With $I=14.90$ and its very high proper motion, 
DENIS~0334$-$49 has a reduced proper motion well above the maximum 
reduced proper motion for an M giant of the same color 
(\citealt{phan-bao03}). This makes it a certain dwarf, and a nearby one.
From an $M_{J}$ vs. $I-J$ relation, calibrated
by 48~single M ($\geq$ M8.0), L, and T dwarfs with known trigonometric 
parallaxes and good photometry (Phan-Bao et al., in preparation),
we estimated a preliminary distance of 5.5$\pm$0.8~pc.

\begin{figure*}
\psfig{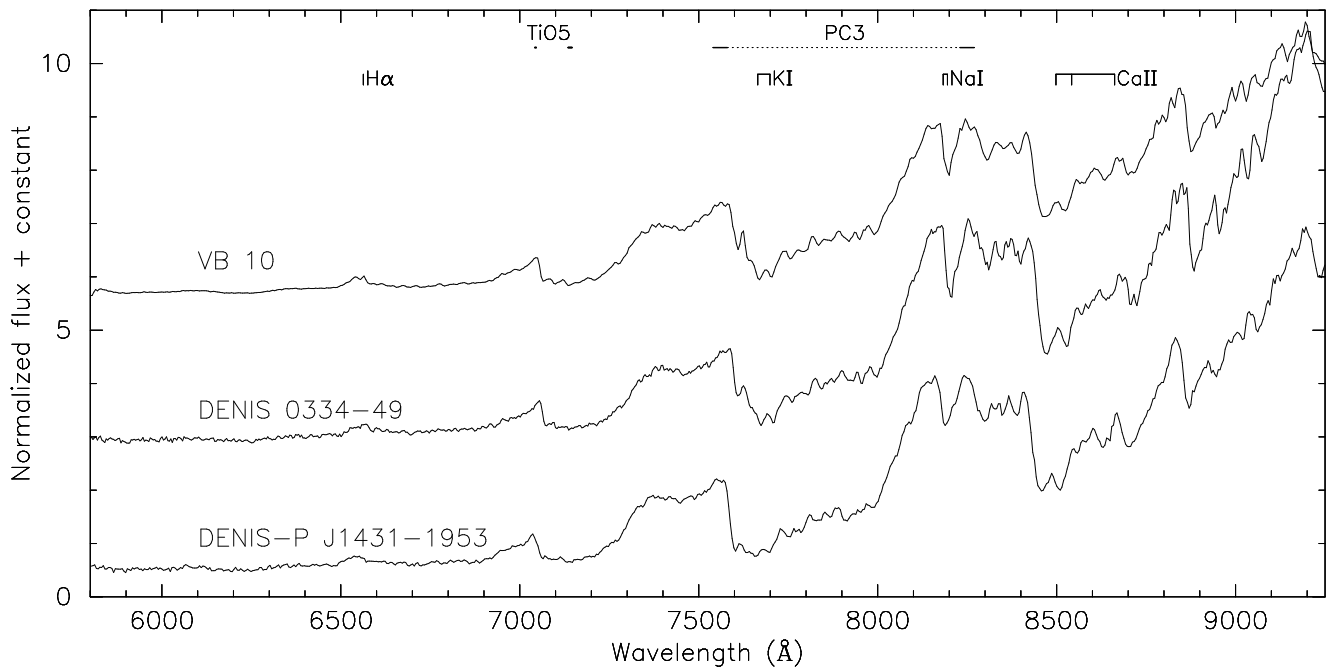}
\caption{Spectrum of DENIS 0334$-$49 (M9, this paper), VB10 (M8, \citealt*{kirkpatrick}); and DENIS-P~J1431$-$1953 
(M9, \citealt{martin99}). The positions of 
the H$_{\alpha}$,  Na{\small I}, K{\small I} and Ca{\small II} lines 
are indicated, as well as the spectral intervals used to compute the 
TiO5, and PC3 indices. 
}
\label{fig_spectra}
\end{figure*}

We observed DENIS~0334$-$49 in September 2005 with the DBS 
spectrograph on the 2.3m telescope at Siding Spring Observatory. The 
158g/mm grating provided a wavelength coverage of 580--1030~nm at 0.5~nm 
resolution. The data were reduced using FIGARO. Smooth spectrum stars were 
observed at a range of airmass to remove the telluric lines using the technique
of \citet{bessell99}. The EG131 \citep{bessell99} spectrophotometric standard 
was used for relative flux calibration, and a NeAr arc provided the 
wavelength calibration. All spectra were normalized over the 754--758~nm 
interval, the denominator of the PC3 index \citet{martin99} and a region 
with a good flat pseudo-continuum. Figure~\ref{fig_spectra} shows the 
resulting spectrum of DENIS~0334$-$49, together with two comparison stars.
The M8.0 standard (VB~10) was observed in the same configuration, and the
M9.0 dwarf (DENIS-P~J1431$-$1953) was observed by \citet{martin99} with
a similar resolution. It is clear from the figure that the spectrum of 
DENIS 0334$-$49 is steeper than that of VB10.

At the resolution of the spectrum, M dwarfs are immediately distinguished
from M giants by the presence of the Na{\small I} and K{\small I} doublets, 
the presence of FeH bands, the appearance of strong CaH cutting into the 
continuum shortward of 700~nm, and by the absence of the Ca{\small II} triplet 
(e.g. \citealt{bessell91}).
\begin{table*}
   \caption{Estimated absolute magnitude, spectrophotometric distance for 
   DENIS 0334$-$49 and VB 10}
    \label{results}
  $$
   \begin{tabular}{llllllllllllll}
   \hline 
   \hline
   \noalign{\smallskip}
Stars    &  PC3   &  TiO5  & Sp.T   & Sp.T    & Sp.T       &    $M_{I}$ & $M_{J}$ & $M_{K}$ & $d_{I}$  & $d_{J}$ & $d_{K}$  & $d_{sp}$   \\
         &        &        &  (PC3) &  (TiO5)  & (adopted) &            &    &   & (pc)  & (pc)  & (pc) &   (pc)                         \\
  (1)     &(2)      &(3)        &(4)  &(5) &(6) &(7) &(8)  &(9) & (10)  & (11)  & (12)  & (13)  \\
   \hline
   \noalign{\smallskip}  
DENIS 0334$-$49&  2.41  &  0.416 &  M9.8 &  M8.6 &  M9.0  &   15.23  &  11.92  &  10.71  &   8.6  &   7.5  &   8.4  &   8.2  \\
VB 10         &  1.87  &  0.307 &  M8.1    &  M8.0 &  M8.0  &  14.29$^{\rm b}$&11.24$^{\rm b}$&10.10$^{\rm b}$& 5.1  &   5.5  & 5.5  & 5.4$^{\rm a}$ \\
    \noalign{\smallskip}
    \hline 
   \end{tabular}
  $$
  \begin{list}{}{}
  \item[$^{\rm a}$]: d$_{\pi}$ = 5.87~pc, derived from $\pi$ = 170.3~mas for
   its proper motion companion HIP 94761
  \item[$^{\rm b}$]: optical and infrared photometry from \citet{bessell91}

{\it Column 1}: Star name.
{\it Columns 2 \& 3}: Spectroscopic indices. PC3 defined in \citet{martin99} and TiO5 in \citet*{reid95}.
{\it Columns 4, 5 \& 6}: Spectral types derived from the PC3 and TiO5 index using the formula given
in \citet{martin99} and \citet{cruz02}.
{\it Columns 7, 8 \& 9}:  Absolute magnitudes for the $I$, $J$, $K$ bands 
based on the PC3-absolute magnitudes relation.
{\it Columns 10, 11 \& 12}: Distance (pc) estimated from the DENIS photometry and
the $M_{\rm I}$, $M_{\rm J}$, $M_{\rm K}$ derived from the PC3 index.
{\it Column 13}: Adopted distance.

  \end{list}
\end{table*}
\section{Discussion}

Table~\ref{results} lists our spectral type estimates for
DENIS~0334$-$49, based on the \citet{martin99} calibration of the PC3 
index and the \citet{cruz02} calibration of the TiO5 index. Since the
TiO5 index wraps around at spectral type $\sim$M7, we used the spectral 
type derived from the PC3 index to choose between the two branches
of the \citet{cruz02} calibration. The VOa index saturates before the
spectral type of DENIS~0334$-$49 and we therefore did not use it. We 
average the spectral types computed from the useful two indices to adopt 
a classification of M9.0$\pm$0.5, consistent as well with visual comparison
with the classification standards. 
H$_{\alpha}$ is not detected, and we used the 
SPLOT IRAF task to measure an upper limit of 2.5~\AA~to its equivalent 
width. Given the weak continuum at  656~nm, this represents a strong
limit on H$_{\alpha}$ emission.

\begin{figure}
\hspace{0.5cm}
\psfig{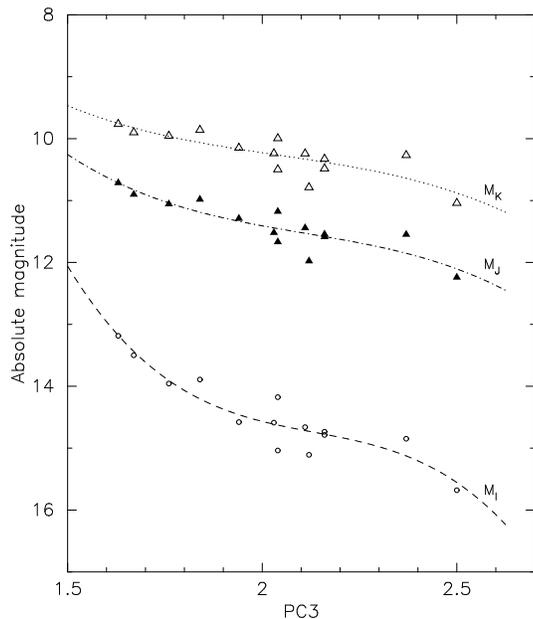}
\caption{The PC3 index vs. absolute magnitudes in the Cousins $I$ and CIT $JK$ passbands
for 14 M dwarfs with spectral types ranging from M7 to M9.5, see Sec.~3.}
\label{fig_PC3}
\end{figure}

To estimate the distance of DENIS~0334$-$49 we have 
extended the \citet{crifo} PC3 vs. absolute magnitude calibration to
higher PC3 index values, adding LP~944-20 and BRI~0021-0214 (respectively
M9.0 and M9.5, and both with data from \citealt{dahn02} and \citealt{geballe})
to their 12 stars later than M7.0. Fig.~\ref{fig_PC3} shows the resulting
PC3 to absolute magnitude relations for the $I$, $J$, and $K$ bands.
The following cubic least-square fits to those data:
\begin{eqnarray}
M_{\rm I} & = & -61.747 + 105.214(PC3) - 48.773(PC3)^{2} \nonumber \\
           &   & + 7.622(PC3)^{3} \label{eq1} \\
M_{\rm J} & = & -16.549 + 37.791(PC3) - 17.404(PC3)^{2} \nonumber \\
           &   & + 2.749(PC3)^{3} \label{eq2} \\ 
M_{\rm K} & = & -7.543 + 24.336(PC3) - 11.479(PC3)^{2} \nonumber \\
           &   & + 1.877(PC3)^{3} \label{eq4} 
\end{eqnarray}

are valid for $1.63 \leq PC3 \leq 2.50$, or spectral types between M7.0
and M9.5. Over this range the rms dispersion of the data around
these fits is approximately 0.2~magnitude, corresponding to a 10\% 
standard error on distances to single stars.

Table~\ref{results} lists the absolute magnitudes for the three DENIS 
bands computed from the PC3 index, as well as the corresponding 
spectrophotometric distance estimates and their average (8.2$\pm$0.8~pc). 
The values for the three colours $I$, $J$, $K$ are very similar, indicating 
correlated uncertainties for the three
estimators. As usual, these distances would be underestimated by 
up to $\sqrt{2}$ if DENIS~0334$-$49 is in fact an unresolved binary.

DENIS~0334$-$49 has a redder $I-J$ color, $I-J=3.59$, than the 
$I-J \sim 3.3$ of a typical M9.0 dwarf, (\citealt{leggett92}). Comparison 
with the DENIS color of well known M9 dwarfs (Table~1 of 
\citealt{phan-bao03}), shows that DENIS~0334$-$49 is much redder at $I-J$ 
than DENIS~1048$-$39 but more similar to LP~944-20 (M9, $I-J=3.27$) or 
BRI~0021-02 (M9.5, $I-J=3.26$); both of these are young.
Additionally, with a high PC3 index 
the DENIS~0334$-$49 absolute magnitudes estimated from the calibration as given above
are $\sim$0.4 magnitude fainter than that of a typical M9 dwarf but consistent with 
that of a young M9 field brown dwarf; raising the 
possibility that DENIS~0334$-$49 might be a brown dwarf and suggesting that a 
lithium test \citep*{martin94} will be of interest. If this is the case,
its distance of 8.2~pc makes DENIS~0334$-$49 the 
3$^{rd}$ nearest M9.0 dwarf in the immediate solar neighbourhood, after 
LP~944-20 (5~pc) and DENIS~1048$-$39 (5.2~pc), and formally before 
LHS~2065 (8.5~pc). 
One should note that in the case of an old M9 field dwarf ($M_{\rm J}=11.45$, \citealt{dahn02})
a derived distance of 9.4~pc would place DENIS~0334$-$49
the 5$^{th}$ closest M9.0, after LHS 1070C ($d=8.8$~pc, \citealt{leinert}).
It is an obvious target for a trigonometric parallax 
measurement, and a good benchmark ultracool dwarf.

\section*{Acknowledgments}
This research is carried out based on the DENIS photometry kindly provided 
by the DENIS consortium. We thank the referee for useful comments that 
improved the paper. This research has made use of the SIMBAD and VIZIER 
databases, operated at CDS, Strasbourg, France.

\label{lastpage}

\end{document}